\documentclass[preprint,aip,jap]{revtex4-1}

\usepackage{graphicx,color}
\usepackage{dcolumn}
\usepackage{bm}
\usepackage{verbatim}
\usepackage{enumerate}
\usepackage{textcomp}
\usepackage[official]{eurosym}
\usepackage{textcomp}
\usepackage[squaren]{SIunits}
\usepackage{amsmath,amssymb}

\newcommand{\lt}{l_{\mathrm{tr}}}
\newcommand{\la}{l_{\mathrm{abs}}}
\newcommand{\zeo}{z_{e_{\mathrm{1}}}}
\newcommand{\zet}{z_{e_{\mathrm{2}}}}
\newcommand{\ze}{z_{e}}
\newcommand{\zp}{z_{p}}

\newcommand{\Tr}{T_{\mathrm{rel}}(\lambda)}
\newcommand{\Tee}{T_{\mathrm{em}}(\lambda)}
\newcommand{\Tt}{T_{\mathrm{tot}}(\lambda)}
\newcommand{\It}{I_{\mathrm{tot}}(\lambda)}
\newcommand{\Io}{I_{\mathrm{0}}(\lambda)}
\newcommand{\Ie}{I_{\mathrm{em}}(\lambda)}
\newcommand{\Y}{\mathrm{YAG:Ce}^{+3}}

\newcommand{\lal}{\lambda_{\mathrm{l}}}
\newcommand{\lar}{\lambda_{\mathrm{r}}}
\newcommand{\lai}{\lambda_{\mathrm{i}}}
\newcommand{\ma}{\mu_{\mathrm{a}}}
\newcommand{\T}{T(\lambda)}
\newcommand{\FIGfull}[3]{
	\begin{figure*}
		\includegraphics{#2}
		\caption{\label{#1} #3}
	\end{figure*}
}

\newcommand{\FIGtt}[3]{
	\begin{figure}[b]
		\includegraphics{#2}
		\caption{\label{#1} #3}
	\end{figure}
}

\newcommand{\FIGwidthe}[4]{
	\begin{figure}[t!]
		\centering
		\includegraphics[width=#4]{#2}
		\caption{\label{#1} #3}
	\end{figure}
}

\begin{document}


\title{How to distinguish elastically scattered light from Stokes shifted light for solid-state lighting?}

\author{M. Meretska}
\affiliation{%
 Complex Photonic Systems (COPS), MESA+ Institute for Nanotechnology, University of Twente, P. O. Box 217, Enschede 7500 AE, The Netherlands\\
}%

\author{A. Lagendijk}
\affiliation{%
 Complex Photonic Systems (COPS), MESA+ Institute for Nanotechnology, University of Twente, P. O. Box 217, Enschede 7500 AE, The Netherlands\\
}%
\author{H. Thyrrestrup}
\affiliation{%
 Complex Photonic Systems (COPS), MESA+ Institute for Nanotechnology, University of Twente, P. O. Box 217, Enschede 7500 AE, The Netherlands\\
}%

\author{A. P. Mosk}
\affiliation{%
 Complex Photonic Systems (COPS), MESA+ Institute for Nanotechnology, University of Twente, P. O. Box 217, Enschede 7500 AE, The Netherlands\\
}%
\author{W. L. IJzerman}
\affiliation{%
 Philips Lighting, High Tech Campus 44, Eindhoven 5656 AE, The Netherlands
}%
\author{W. L. Vos}%

\affiliation{%
 Complex Photonic Systems (COPS), MESA+ Institute for Nanotechnology, University of Twente, P. O. Box 217, Enschede 7500 AE, The Netherlands\\
}%

\date{\today}

\begin{abstract}
We have studied the transport of light through phosphor diffuser plates that are used in commercial solid-state lighting modules (Fortimo). These polymer plates contain $\Y$ phosphor particles that elastically scatter light and Stokes shifts it in the visible wavelength range (400-700 nm). We excite the phosphor with a narrowband light source, and measure spectra of the outgoing light. The Stokes shifted light is separated from the elastically scattered light in the measured spectra and using this technique we isolate the elastic transmission of the plates. This result allows us to extract the transport mean free path $\lt$ over the full wavelength range by employing diffusion theory. Simultaneously, we determine the absorption mean free path $\la$ in the wavelength range 400 to 530 nm where $\Y$ absorbs. The diffuse absorption $\ma =\frac{1}{\la}$ spectrum is qualitative similar to the absorption coefficient of $\Y$ in powder, with the $\ma$ spectrum being wider than the absorption coefficient. We propose a design rule for the solid-state lighting diffuser plates.

\end{abstract}

\maketitle


\section{\label{ch:introduction}Introduction}

Energy efficient generation of white light is attracting much attention in recent years, since it is important for lighting and for medical and biological applications \cite{Schmidt99,Whelan00,Malakoff02,Eells04,Taguchi04,Schubert06,Krames07,Bechtel08,Breslauer09,Sommer09}. One of the main directions is the technology of solid-state lighting \cite{Schubert06,Krames07,Bechtel08,Sommer09}, that was recognized with the 2014 Nobel Prize in physics~\cite{Aka14}. Solid-state lighting provides superior energy efficiency and flexibility in terms of color temperature. Conventional solid-state lighting employs a blue semiconductor light emitting diode (LED) in combination with a phosphor layer to realize a white-light emitting diode. The phosphor layer plays two important roles in a white LED: first, the phosphor layer absorbs blue light emitted by the LED, and efficiently converts part of the blue light into the additional colors green, yellow and red light. The desired mixture of blue, green, yellow and red light results in white outgoing light. Secondly, the phosphor layer multiply scatters all colors, thereby diffusing the outgoing light, resulting in an even lighting without hot spots, and with a uniform angular color distribution, as required for lighting applications. In addition the scatterers enhance the color conversion by increasing the path blue light travels in phosphor layer. In state-of-the art solid-state lighting technology the phosphor layer is engineered to have a complex internal structure\cite{Bechtel08,Sommer09}. Light inside this layer may be multiply scattered not only by phosphor, but also by other scatterers. 

\FIGwidthe{fig:tr1}{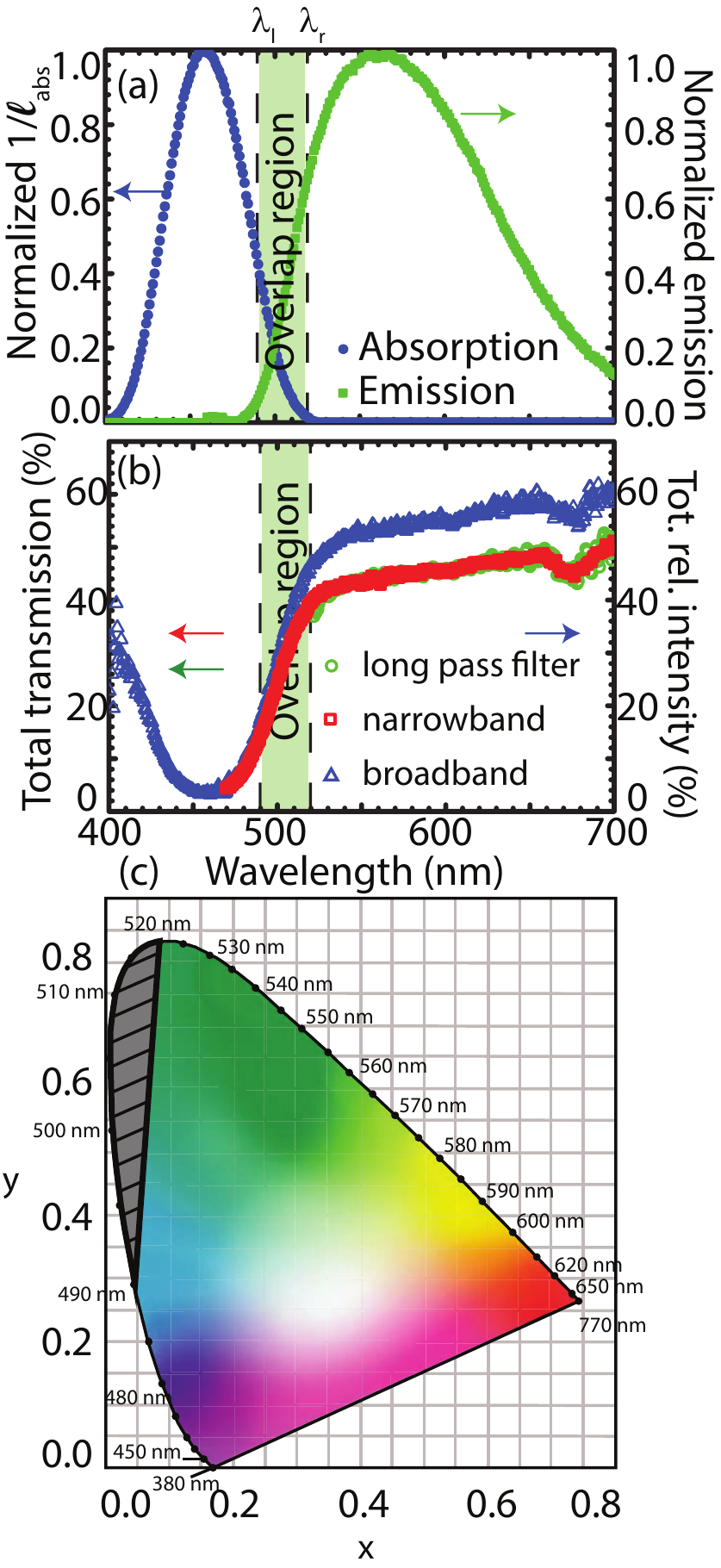}{\textbf{(a)} Normalized absorption spectra (blue circles) and emission spectrum (green squares) of the $\Y$ phosphor used in our study. The spectral range where emission and absorption spectra overlap is indicated with a green bar between $\lal$ and $\lar$. \textbf{(b)} Transmission spectra obtained by using the narrowband (red squares) and the broadband light sources (green circles) for the polymer plate with 4 wt~$\%$ phosphor particles. Arrows point to the relevant ordinate for the data. \textbf{(c)} CIE 1931 (x,y) chromaticity diagram~\cite{Schubert06}. Monochromatic colors are located at the perimeter of the diagram. In the middle of the diagram the white color is located. The dashed gray area represent the region where emission and absorption of $\Y$ overlap. The overlap range was previously inaccessible and it is made accessible in this work.}{3 in}

In spite of the wide use of solid-state lighting in everyday life, and the apparent simplicity of the physical processes occurring in the phosphor layer, there is no analytical theory that predicts the spectra of white LEDs. The main challenge arises from the lack of physical understanding of systems, where multiple scattering and absorption of blue light coexist with emission of light in a broadband wavelength range. Typically, numerical methods such as ray-tracing and Monte Carlo techniques are used \cite{Sommer09,Gilray96,Liu10,Tukker10}, that do not have the predictive power of analytical theory. These simulations require a set of heuristic parameters that is derived from measurements on LEDs with a wide range of structural and optical parameters. The resulting heuristic parameters are often adjusted, thereby further hampering the predictive power of these methods. Moreover, simulations are time consuming and computationally demanding. All these aspects hamper the efficient design of new white LEDs.

In this paper we - for the first time - extract optical properties of the phosphor layers typically used for the solid-state lighting in the visible wavelength range (400-700 nm). We use a narrowband light source and record the spectra of the transmitted light through the phosphor layer. The transmitted light contain both elastically scattered light and Stokes shifted light; they are separated spectrally. We extract the diffuse transmission from the elastically scattered light, and calculate the optical properties of the phosphor layer using the diffusion theory. Using a broadband light source a similar approach was previously applied to calculate optical properties of the diffuser plates~\cite{Vos13}, and phosphor plates~\cite{Leung14}. This approach fails when Stokes shifted light overlaps spectrally with the elastically scattered light. In Fig.~\ref{fig:tr1}(c) we show the overlap region for the plates studied here in $x-y$ chromaticity diagram. This spectral range correspond to the green part of the white LED spectrum where the human eye is most sensitive\cite{Malacara11}. We present a new measurement technique that allow us to separate the elastically scattered and Stokes shifted light in the overlap range for phosphor diffuser plates that are used in commercial white LEDs. As a result we now close the ''the green gap'' and extract the relevant transport and absorption parameters for solid-state lighting in the whole visible spectral range. 

We use analytical theory originating from nanophotonics, wherein propagation of light is described from first principles \cite{Lagendijk96,Rossum99,Akkermans07,Garcia08,Muskens08}. Such $ab$ $initio$ theory supplies fundamental physical insights on the light propagation inside solid-state lighting device~\cite{Vos13,Leung14}. Extracting the optical parameters from theory are less time consuming than performing many simulations, and more importantly, the resulting parameters are robust and predictions can be made beyond the parameter range that was initially studied. For instance, knowledge of the absorption spectra provides us with the design guidelines for the solid-state light units. The design parameters such as the thickness of the diffuse plates and the phosphor concentration can be directly extracted from the absorption spectra depending on the blue pump wavelength of a white LED.

\section{\label{ch:theory}Theory}

\subsection{\label{ch:tr}Total transmission with energy conversion}

Multiple light scattering is usually studied by measuring the total transmission through a slab of a complex, multiple scattering medium~\cite{Hulst80,Kokhanovsky2015}. Total transmission, or diffuse transmission, is the transmission of an incident collimated beam with intensity $\Io $ that is multiple scattered and integrated over all outgoing angles at which light exits from a medium. The total transmission carries information on the transport mean free path $\lt$ and on the absorption mean free path $\la$, which are the crucial parameters that describe multiple light scattering \cite{ Ishimaru78, Garcia92, Durian95, Elaloufi02, Akkermans07,Sarma15}. The transport mean free path $\lt$ is the distance it takes for the direction of light to become randomized while performing a random walk in a scattering medium.  The absorption mean free  path $\la$ is the distance it takes for light to be absorbed to a fraction $(1/e)$ while light performs a random walk in a scattering medium.

Phosphor particles do not only scatter light, but also convert blue light into other colors by absorbing blue and re-emitting other colors of light. Therefore, from here on we will refer to the measured total transmission in presence of energy conversion as the $total$ $relative$ $intensity$ $\Tr$
\begin{equation}
\Tr =\frac{\It }{\Io}\mathrm{,}
\label{eq:trel}
\end{equation} 
where $\It$ is the integrated intensity that is collected at the back side of the diffusion plate. In the emission range of a phosphor $(\lambda\geq\lal)$ the collected intensity $\It$ can be written as a sum of the diffuse intensity $I$ and the Stokes shifted intensity $\Ie$.  
Thus, the total relative intensity can be separated into two parts~\cite{Leung14}
 \begin{equation}
\Tr=\T+\Tee=\frac{I (\lambda)+\Ie}{\Io}\mathrm{,}
\label{eq:trel1}
\end{equation} 
where the first term $T$ is the total transmission, and the second term $\Tee $ is the emission that accounts for the energy conversion of light in the diffuse absorptive medium.        
In Ref.~\onlinecite{Leung14} these two terms could not be distinguished in the overlap range $\lal < \lambda < \lar$.

The central question in this paper is how to distinguish the total transmission $\T$ from the total relative intensity $\Tr$ as this allows one to obtain both the transport mean free path $\lt$ and the absorption mean free path $\la$.
To access the total transmission we employ a tunable narrowband light source and spectrally resolve the narrowband transmitted light. Since the light that is converted by the phosphor exhibits a Stokes shift $\Ie$, this part $\Tee$ of the total relative intensity is filtered, hence we obtain the desired total transmission $\T$ that we interpret with diffusion theory.

\subsection{Total transmission in absence of energy conversion}

According to diffusion theory for light, the total transmission $\T$ through a slab, even in the presence of absorption, is a function of the slab thickness $L$, the wavelength $\lambda$, the transport mean free path $\lt$, and the absorption $\la$ mean free path, and can be expressed as \cite{Ishimaru78}~\footnotemark[1]:

\begin{equation}
T\left( L, \lai , \lt, \la\right) =Q^{-1}\left[ \mathrm{sinh}\left( \ma z_{p}\right) +\ma\ze \mathrm{cosh}\left(\ma z_{p} \right) \right] \mathrm{,}
\label{eq:tr}
\end{equation} 
with
\begin{equation}
Q\left( L, \lai , \lt, \la\right)\equiv \left( 1+ \ma^{2}\ze^{2}\right) \mathrm{sinh}\left( \ma L\right) +2\ma\ze \mathrm{cosh}\left(\ma L \right) \mathrm{,}
\label{eq:ex}
\end{equation}
where the extrapolation lengths are equal to
\begin{equation}
\zeo \left(\lt\right)=\zet \left(\lt\right)=\ze \left(\lt\right)= \frac{2}{3} \lt \frac{1+\overline{R_{1,2}}}{1-\overline{R_{1,2}}}\mathrm{.}
\label{eq:ex1}
\end{equation}
Here $\zp$ is the diffuse penetration depth of light, $\ma \equiv 1/\la$ the inverse absorption mean free path, $\overline{R_{1,2}}$ is the angular and polarization averaged reflectivity of the respective boundaries \cite{Lagendijk89}. For a normal incident collimated beam the penetration depth becomes $\zp=\lt$ \cite{Gomez99}, and $\overline{R_{1,2}}$=0.57 for polymer plates with an average refractive index $n=1.5$ \cite{Zhu91}. 
For samples with no absorption ($\ma = 0$), Eq.~\eqref{eq:tr} simplifies to the optical Ohm's law~\cite{Lagendijk96}
\begin{equation}
T\left( L, \lai , \lt\right)=\frac{\lt +\ze}{L+2\ze}\mathrm{.}
\label{eq:tr1}
\end{equation} 

In the range of zero phosphor absorption the total transmission is a function of the sample thickness $L$, the incoming wavelength $\lai$, and the transport mean free path $\lt$, $T=T(L,\lai , \lt)$. Therefore, we can extract $\lt$ using Eq.~\eqref{eq:tr1} from measurements of the total transmission $\T$ in the range of no absorption $(\lai\geq \lar)$. In the range of strong phosphor absorption $(\lai\leq\lar)$ the total transmission also depends on the absorption mean free path $\la$: $T=T(L,\lai , \lt, \la)$. Therefore the transport mean free path $\lt$ has to be derived separately, which we can do by extrapolating the $\lt$ values extracted in the zero absorption wavelength range $(\lai\geq\lar)$ into the strong absorption wavelength range, since $\lt$ is a monotoneously increasing function of $\lambda$ for size-polydisperse scatterers~\cite{Muskens09,Muskens08,Vos13}. By measuring the total transmission in the range of strong absorption, we thus obtain $\la$ using extrapolated values of $\lt$, by using Eq.~\eqref{eq:tr}.

\section{\label{ch:Exp}Experimental details}

We have studied the light transport through polymer plates that are used in Fortimo solid-state lighting units \cite{LED14}. The polymer plates consist of a polycarbonate matrix (Lexan 143R) with  $\Y$ ceramic phosphor particles that are widely used in white LEDs. The phosphor particles have a broad size distribution with center around 10$\ \mu$m \cite{Leung14}, and a $\mathrm{Ce^{3+}}$ concentration in the $\Y$ of 3.3 wt $\%$.

The emission and absorption spectra of the $\Y$ in powder form, which was used in the polymer diffusion plates are shown in Fig.~\ref{fig:tr1}(a). The absorption and emission bands have peaks at 458~nm and 557~nm, respectively, and overlap in the spectral range between $\lal$=490~nm and $\lar$=520~nm. As a result, we distinguish three spectral ranges in the visible spectrum where different physical processes are taking place: (1) in the spectral range up to $\lal=$490~nm, light is partly elastically scattered and partly absorbed. This range has already been studied in Ref.~\onlinecite{Leung14}. (2) In the spectral range between $\lal=$490~nm and $\lar=$520~nm, externally incident light is elastically scattered and absorbed, while light is also emitted by the internal phosphor, and subsequently elastically scattered. This is the overlap range that is central to the present work. (3) In the spectral range beyond $\lar=$520~nm, output light is the sum of externally incident light that is elastically scattered and of internally emitted light by the phosphor that is also elastically scattered. This range has also already been studied in Ref.~\onlinecite{Leung14}.

Here we present the transmission measurements on five such polymer plates with a phosphor concentration ranging from $\phi=2.0$ to $4.0$ wt$\%$. The corresponding volume fractions range from $\phi=0.5$ to $1$ vol$\%$, which is in the limit of low scatterer concentration. The plates were prepared using injection molding where a powder of $\Y$ particles is mixed with the polymer powder and the mixture is melted and pressed into a press-form. The polymer plates, shown in Fig.~\ref{fig:set}(a), are circular with a diameter of 60 mm and a thickness of 2 mm.

Figure~\ref{fig:set}(b) shows a drawing of the setup for measuring the spectrally resolved total transmission $\Tt$ with a narrowband incident light beam. We illuminate the sample with two different light sources: a tunable narrowband light source and a broadband light source. The narrowband light source consists of a Fianium supercontinuum white-light source (WL-SC-UV-3), which is spectrally filtered to a bandwidth of less than $\Delta\lambda=$2.4 nm by a prism monochromator (Carl-Leiss Berlin-Steglitz). The wavelength of the narrowband source is tunable in the wavelength range between 400 and 700 nm, as shown in Fig~\ref{fig:sp}. 
The infrared part above 700 nm of the supercontinuum laser source is filtered  by a neutral density filter (NENIR30A) and a dichroic mirror (DMSP805). The spectrum of the supercontinuum source after filtering the infrared light is shown in Fig.~\ref{fig:sp}. The spectrum vary drastically in intensity at different wavelengths.

The incident beam illuminates the phosphor plate at normal incidence and the plate is placed at the entrance port of an integrating sphere. We verified that the entrance port of the integrating sphere is sufficiently large to collect all intensities that emanated from the strongest scattering sample. The intensity of the outgoing light entering the integrating sphere is analyzed with a fiber-to-chip spectrometer (AvaSpec-USB2-ULS2048L) with a spectral resolution of $\Delta\lambda=$2.4 nm. 

An example of the measured spectra for three incident wavelengths $\lai$=527, 550 and 634 nm are shown in Fig.~\ref{fig:sp} with the red squares and red dotted lines. The peak intensity varies significantly across the spectral range as a result of the spectral variation of the supercontinuum source. The integration time is changed at every wavelength during measurements to maintain a fixed reference intensity. 

The alternative broadband light source consist of a white LED (Luxeon LXHL-MW1D) (see Fig.~\ref{fig:set}) with an emission spectrum covering the range from 400 nm to 700 nm as shown in Fig.~\ref{fig:sp}. This source is not filtered.

\FIGfull{fig:set}{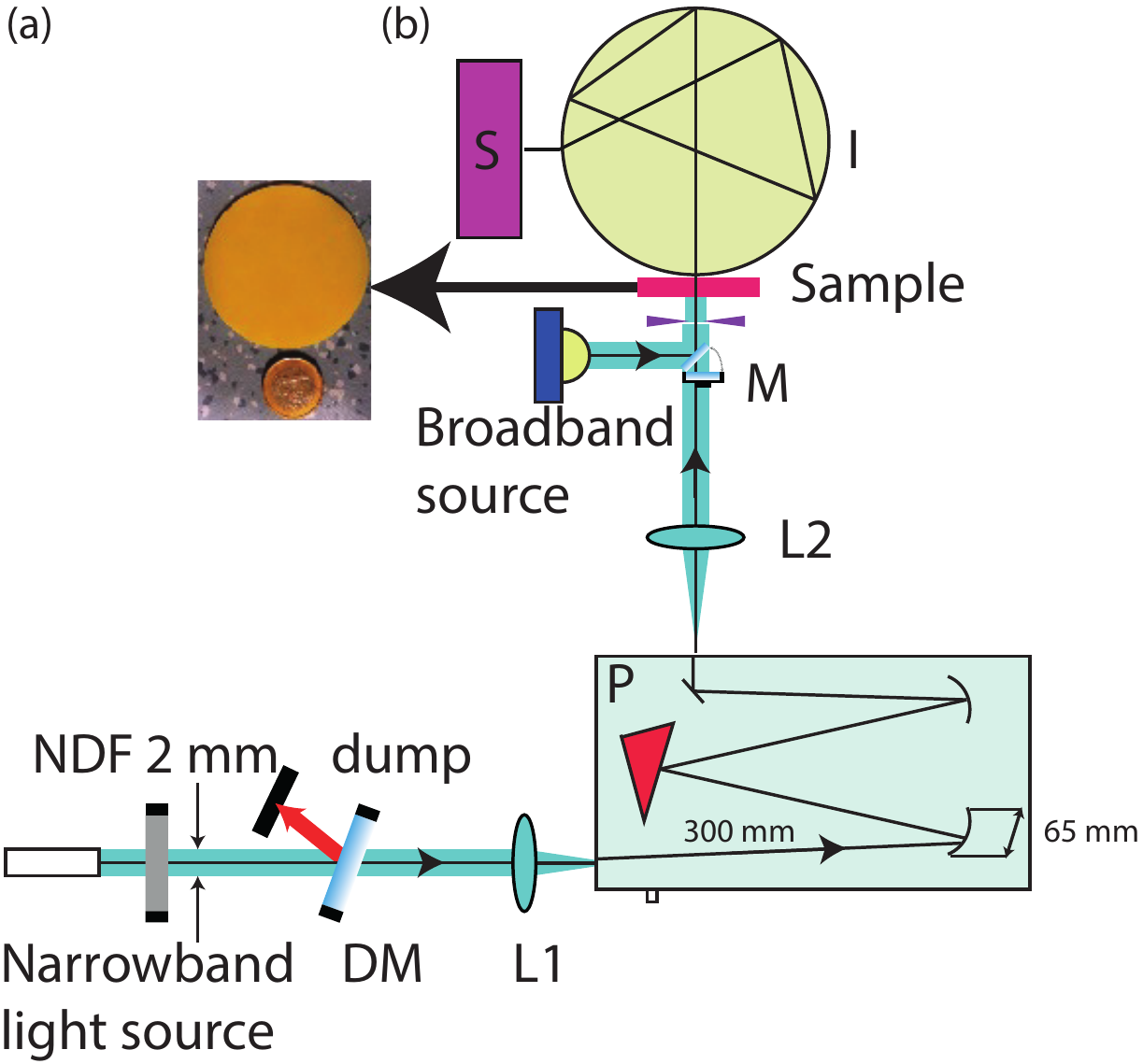}{Narrowband measurement setup. \textbf{(a)} A polymer slab with a 4 wt$\%$ $\Y$ compared to a $\texteuro{1}$ coin. \textbf{(b)} Supercontinuum white light source Fianium, NDF: neutral density filter, DM: dichroic mirror, L1: achromatic doublet (AC080-010-A-ML, f=10 mm), L2: achromatic doublet (f=50 mm), M: mirror, I: integrating sphere, S: spectrometer, P: prism spectrometer ($f_{\sharp} = 4.6$). }

\FIGtt{fig:sp}{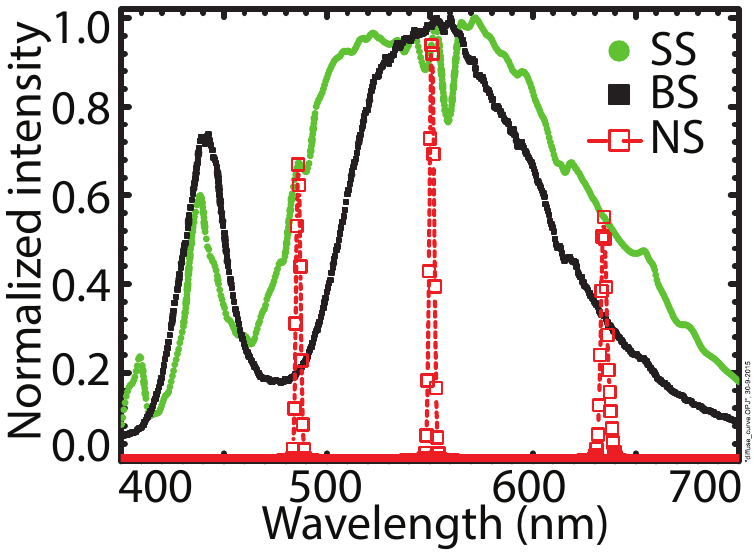}{Normalized reference spectra of the light sources used in the experiment. Blue circles - normalized reference spectrum of the supercontinuum source (SS) after being filtered by a neutral density filter and a dichroic mirror that removes the infrared part of the spectra. Black squares - normalized reference spectrum of a broadband source (BS) that was not filtered. Red squares - spectrally filtered narrowband light source (NS) normalized to the initial BS spectrum represent. We show narrowband spectra for three incident wavelengths $\lai=486,551,634$~nm. Intensity normalized to the SS.}

For all phosphor plates, we measured the transmission spectra with both the narrowband and the broadband light source to check the consistency in the spectral range where both methods can be applied ($\lambda\leq\lal$ and $\lambda\geq\lar$). 
For the broadband light source the total transmission is obtained by normalizing the measured spectrum $I(\lambda )$ to a reference spectrum $\Io$ measured in the absence of a sample. For the narrowband light source the total transmission is determined as the ratio of the transmitted intensity $I(\lai )$ and a reference intensity $I_0 (\lai )$ without the sample at the designated wavelength $\lai$ that is set with the monochromator. The total transmission is reproducible to within a few percent points on different measurements with different light sources.

\section{\label{ch:Results}Results}
\subsection{Transmission measurements}

 In order to measure the total transmission $\T$ of the polymer plates we tune the narrowband light source to an incident wavelength $\lai$, and measure the transmitted intensity of the outgoing light. In Fig.~\ref{fig:em} we show the normalized transmitted intensity for three incident wavelengths $\lai$. For an incident wavelength $\lai$=490 nm, we see a pronounced peak with the maximum at $\lambda$=490 nm. This peak contains mostly elastically scattered photons, because inelastically scattered photons are Stokes shifted to longer wavelengths. Indeed between 500 nm and 650 nm the intensity profile reveals a broad peak that represents the Stokes shifted intensity, since the intensity profile has a shape similar to that of the $\Y$ in powder form (black dashed curve). Both emission spectra have a maximum at 557 nm. For $\Y$ in powder form the intensity is slightly higher at longer wavelengths $\lambda > 600$~nm than for the phosphor plate. 
 The peak value of the Stokes shifted light amounts to 3 $\%$ of the transmitted intensity at the incident wavelength $\lai$=490 nm. We calculated the amount of Stokes shifted light $\Ie$ in the elastic peak at $\lai$=490 nm by extrapolating the emission curve to this wavelength range. We find that $\Ie$ amounts to less than 1 $\%$ of the transmitted intensity I($\lambda$), and can be neglected safely. The total Stokes shifted intensity decreases drastically for longer incident wavelengths $\lai$ inside the overlap range, and can thus be neglected too. For incident light in the middle of the overlap range at $\lai$=505 nm, we see that the emitted intensity is even weaker with a normalized intensity $\Ie$ less than 1$\%$ of I($\lambda$). At the edge of the absorption band at the incoming wavelength $\lai$=520 nm we observe an even smaller Stokes shifted intensity $\Ie$. The reason for this decrease is that the absorption cross section decreases drastically in the overlap range, and only a very little amount of light is being absorbed, and as a result is Stokes shifted. Throughout the overlap range the contribution from the Stokes shifted light $\Ie$ is less then 1~$\%$ of I($\lambda$) at the incident wavelength $\lai$. We thus conclude that the Stokes shifted intensity contribution can be neglected throughout the overlap range. Therefore, we have distinguished the elastically scattered (or absorbed) fraction of light from the Stokes shifted light. As a result we can now measure total transmission at any desired wavelength.

We have scanned the incident wavelength $\lai$ through the wavelength range of interest. In Fig.~\ref{fig:tr1}(c) the total transmission for the slab with 4 wt$\%$ of $\Y$ has been obtained from these scans (red). The total relative intensity $\Tr$ measured with the broadband source is also shown in Fig.~\ref{fig:tr1}(c) for comparison (blue). For short wavelengths, both transmission spectra coincide within a few percent. The blue spectrum reveals a deep trough with a minimum at 458 nm. The trough matches well with the peak of the absorption band of $\Y$ in Fig.~\ref{fig:tr1}(a), and reveals that a significant fraction of the light in this wavelength range is absorbed by the phosphor. At wavelengths longer than $\lar$=520 nm both transmission spectra are flat, but the spectrum measured with the broad band light source is 10 $\%$ larger than the spectrum measured with the narrowband light source. The blue triangle spectrum contains a significant contribution of the Stokes shifted light $\Ie$ in this spectral range, as most of the emission occurs in this spectral range (see Fig.~\ref{fig:tr1}(a)). The narrowband spectrum on the contrary does not have this contribution. The difference in transmission between these two spectra at long wavelengths is equal to $\Tee$, in Eq.~\eqref{eq:trel1}. In the overlap region ($\lal=490$~nm$<\lambda<\lar=520$~nm) both spectra reveal a sharp rise. The total relative intensity spectrum deviates from the total transmission spectrum due to the contribution of the Stokes shifted light $\Ie$ in this wavelength range. In Ref.~\onlinecite{Leung14} it was not possible to separate elastically scattered and Stokes shifted light.

 Finally, we filtered the broadband light source with a longpass filter at $\lambda$=520 nm, which ensures that the phosphor is not excited (we thus have zero emitted intensity $\Tee$=0). Hence, the measured total relative intensity equals the total transmission $\Tr=\T$ in this spectral range. In Fig.~\ref{fig:tr1}(b) we compare the total transmission measured with the filtered broadband light source (green), and the total transmission measured with the narrowband light source (red). The total transmission measured with the narrowband light source agrees within a few percent points with the total transmission measured with the filtered broadband light source (see Fig.~\ref{fig:tr1}(c)) for $\lai\geq\lar$. In summary, we have for the first time extracted the total transmission $\T$ for a diffuser plate with phosphor in the whole visible range, including the previously inaccessible~\cite{Leung14} overlap range.

\FIGtt{fig:em}{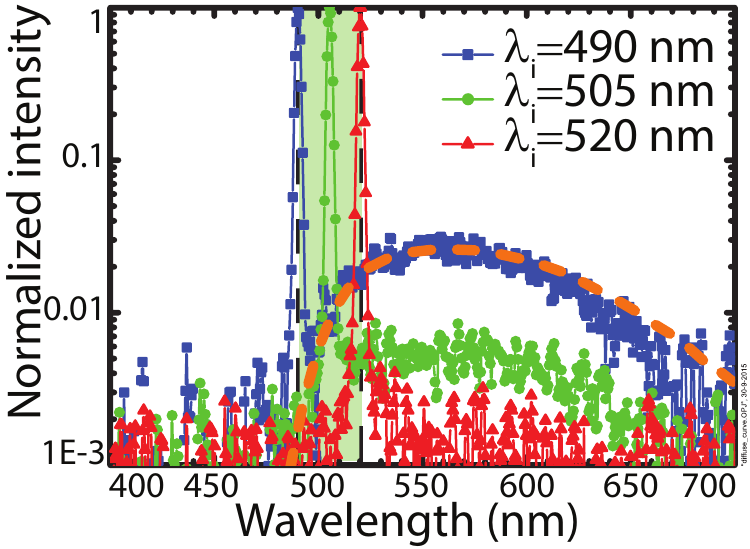}{Intensity profile of the signal that we measure in the range where emission and absorption overlap for three different pump wavelengths. The orange dashed line is the emission spectra from Fig.~\ref{fig:tr1}(a). The blue curve was normalized to 7791 counts and the green and the red curves to 15480 and 20297 counts respectively.}

\subsection{Transport mean free path}

\FIGfull{fig:tr2}{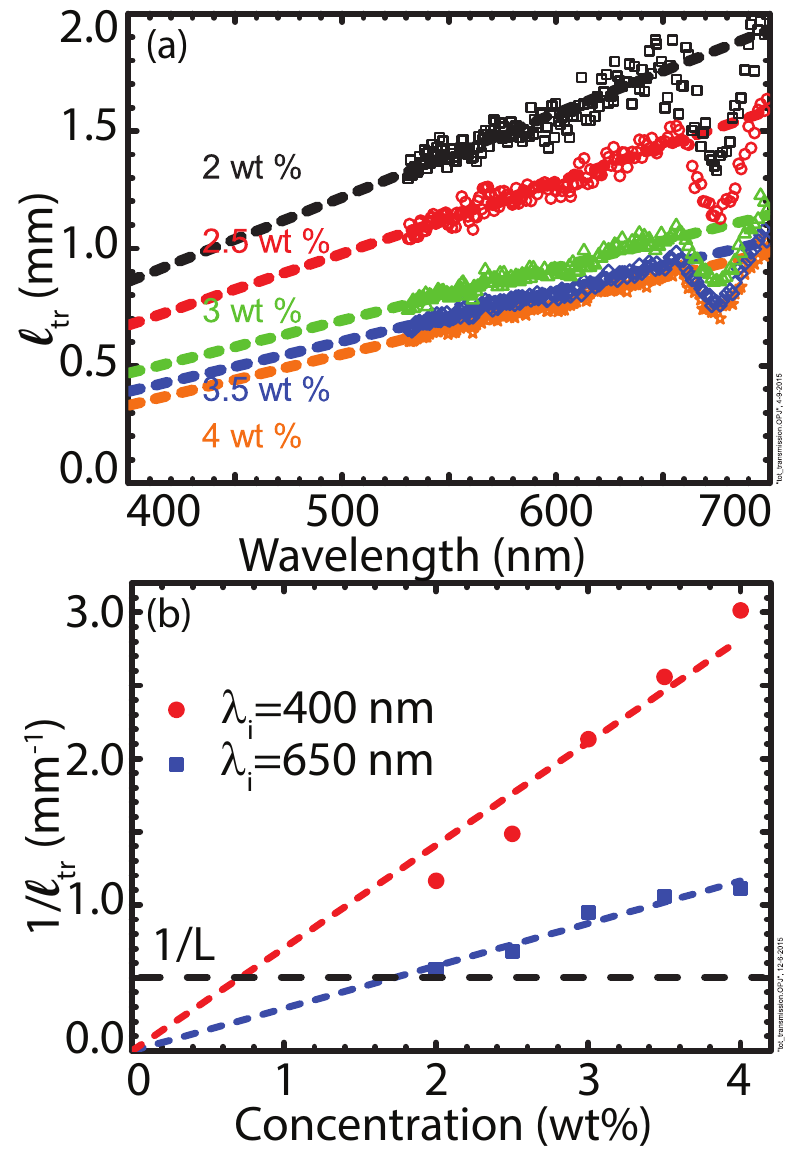}{Transport mean free path. \textbf{(a)} Transport mean free path as a function of wavelength for different phosphor concentrations. Dashed lines represent linear fits to the measured data with parameters listed in Tab.~\ref{tab:f}. \textbf{(b)} Inverse transport mean free path as a function of concentration for two different incoming wavelengths. Dashed lines are linear fits to the data.}

In the range of low absorption we have extracted the transport mean free path from the transmission data using Eq.~\ref{eq:tr1} and plotted the result in Fig.~\ref{fig:tr2}(a). We see that the transport mean free path  increases linearly with wavelength at constant phosphor concentration. In highly polydisperse non-absorbing media, a similar relationship between the transport mean free path and wavelength was found and interpreted~\cite{Vos13}. Therefore, we have fitted the transport mean free path with a line for every phosphor concentration. Parameters of the linear fits are listed in appendix. We linearly extrapolate $\lt$ to the absorption range $\lambda\leq\lar=520$~nm, and use the extrapolated values of the transport mean free path $\lt$ to obtain the absorption mean free path $\la$ in the range of strong absorption.

\begin{table}[h!]
\begin{center}

    \begin{tabular}{ | c | c | c | c | c |}
    \hline
    C (wt$\%$) & a & b  \\ \hline
    2 & -0.57 $\pm$ 0.07&  0.00358  $\pm$ 0.00001\\ \hline
    2.5 &-0.55$\pm$ 0.05 & 0.00305 $\pm$ 0.00008\\ \hline
    3 & -0.43$\pm$ 0.03 &0.00225 $\pm$ 0.00055\\ \hline
    3.5 & -0.47$\pm$ 0.02 & 0.00214 $\pm$ 0.00038\\ \hline
    4 & -0.54$\pm$ 0.02 & 0.00218$\pm$ 0.00038 \\ 
    \hline
    
    \end{tabular}
\end{center}
\caption{Parameters of the linear models of the transport mean free path versus wavelength: $\lt = a + b\lambda$. The parameters $a$ and $b$ depend on the phosphor concentration $C$, and are shown with their standard errors.}
\label{tab:f}
\end{table}

In the limit of low concentration each scatterer can be treated independently. In this case $1/\lt$ is proportional to the concentration of the scatterers~\cite{Busch94}. Indeed in Fig.~\ref{fig:tr2}(b) we see that the transport mean free path increases inversely proportional with the increasing phosphor concentration at a fixed wavelength. We observe that with increasing wavelength the scattering cross section increases similarly to what was obtained earlier~\cite{Leung14}.

\FIGfull{fig:trm}{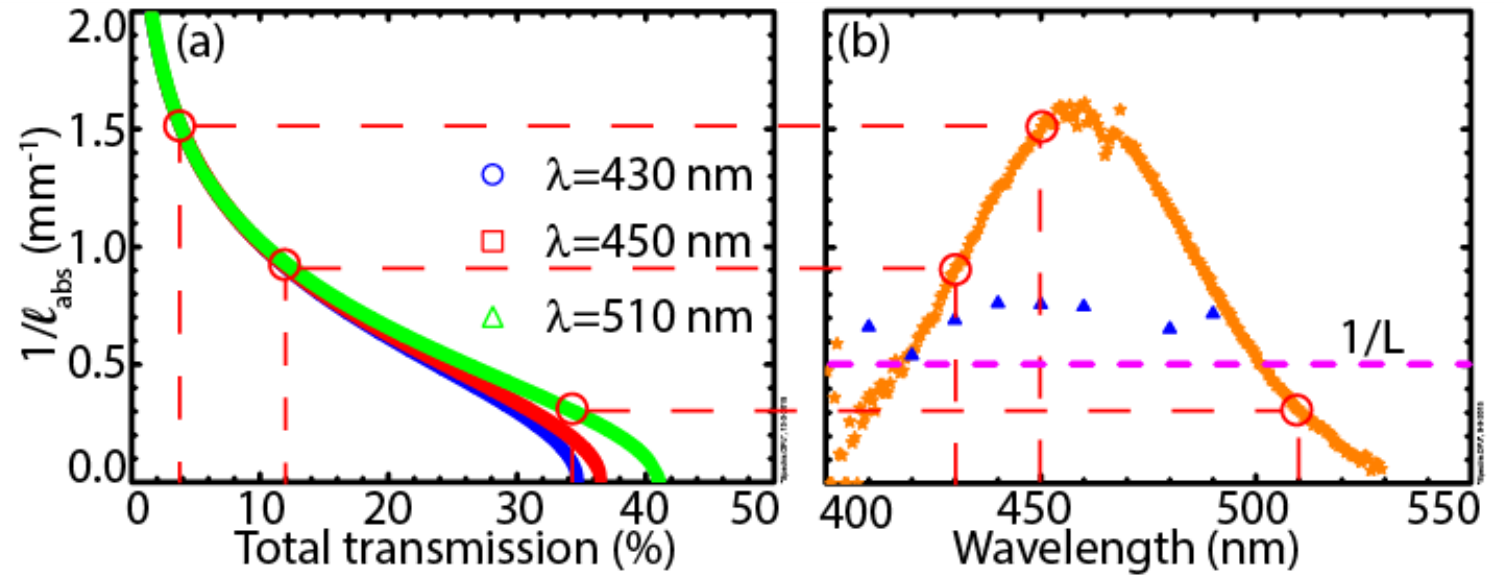}{Determination of the absorption mean free path. \textbf{(a)} Look-up tables presented in a form of plots for the Eq.~\ref{eq:tr} for three different wavelengths, and 4 wt$\%$ concentration of the phosphor. ($\lt$=0.40;0.44;0.57 $\mathrm{mm}$ respectively) \textbf{(b)} Absorption mean free path extracted from the look-up tables for the plate with 4 wt$\%$ phosphor concentration. Red circles and red dashed lines indicate the process of mapping $\ma$ from look-up table to the Fig.~\ref{fig:trm}(b). Blue triangles indicate the inverse absorption length measured in Ref.~\cite{Leung14}}

\subsection{Absorption mean free path}
By using the transport mean free path $\lt$ extrapolated to the wavelength range between 400 and 530 nm where the phosphor absorbs light, we now derive the absorption mean free path $\la$  from the measured total transmission (see Fig.~\ref{fig:tr1}(c)). Since we do not have analytic inverse function of Eq.~\eqref{eq:tr} , we have solved the inversion numerically and made look-up tables for each phosphor concentration and at each wavelength. Fig.~\ref{fig:trm}(a) shows three inverted curves of $\ma$ versus total transmission for three different wavelengths ($\lambda=$430, 450, 510~nm) at a phosphor concentration $C$=4 wt$\%$. We plot $\ma$ - rather then $\la$ - since this quantity tends to zero for vanishing absorption.
Fig.~\ref{fig:trm}(a) shows that $\ma$ (and thus the absorption) increase. In the limit of strong absorption, all total transmission curves tend to zero. In the limit of vanishing $\ma$, the total transmission equals the (extrapolated) values that decrease with decreasing wavelength (Fig.~\ref{fig:trm}(a)). The vertical dashed lines indicate the measured total transmission, and the intersections with the curves yield the corresponding $\ma$ for each wavelength at this phosphor concentration. 
 
 Fig.~\ref{fig:trm}(b) shows the extracted absorption profile $\ma$ for the polymer plate with the highest  phosphor concentration $C$=4 wt$\%$ studied here. The FWHM of this curve is 64.5~nm. The dashed purple line in Fig.~\ref{fig:trm}(b) indicates the inverse thickness of the sample. The absorption mean free path $\la$ is shorter than the thickness of the sample $L$ between 418 and 501 nm. This means that incident light is effectively absorbed in the volume of the sample, and the density of the phosphor is optimized for use in a white-light LED. At the edges of the absorption range at 400 and 530~nm $\ma$ tends to zero, as expected from the known absorption~(Fig.~\ref{fig:tr1}(a)). We note that our present $\ma$ values differ from previous results obtained on the same samples~\cite{Leung14} (see Fig.~\ref{fig:trm}(b)). 
The absorption mean free path varies significantly with wavelength in our case. We notably attribute the difference to the use of an incorrect diffusion equation in Ref.~\onlinecite{Leung14}. The spectral shape obtained at present is in much better agreement with the phosphor absorption spectra than previously, which is gratifying.

In Fig.~\ref{fig:trm1}(a) we have plotted $\ma(\lambda)$ for three wavelength $\lambda$ = 430, 460 and 500 nm as a function of phosphor concentration. We see that $\ma(\lambda)$ increases linearly with increasing phosphor concentration, which agrees with the assumption that each absorber is independent. Figure~\ref{fig:trm1}(a) shows that the steepest slope appears at the wavelength $\lambda$=460 nm, which corresponds to the peak of the absorption curve. The maximum absorption cross section is $\sigma_{\mathrm{abs}}=$30~$\mu m^{2}$, which agrees reasonably well with the typical measured absorption cross section for $\Y$ of the order of 10~$\mu m^{2}$ (Ref.~\onlinecite{Liu10,Kaczmarek99,Zhao03,Mares07,Mihokova07,Kucera08}).

Figure~\ref{fig:trm1}(b) shows the extracted normalized absorption spectrum of $\ma$ for the polymer plates with different phosphor concentrations. All absorption curves are normalized to their maximum, and they coincide with each other within a few percent points. Absorption mean free path $\la$ scales linearly with the phosphor concentration. All absorption curves tend to zero outside $400<\lambda<530$~nm.

\FIGfull{fig:trm1}{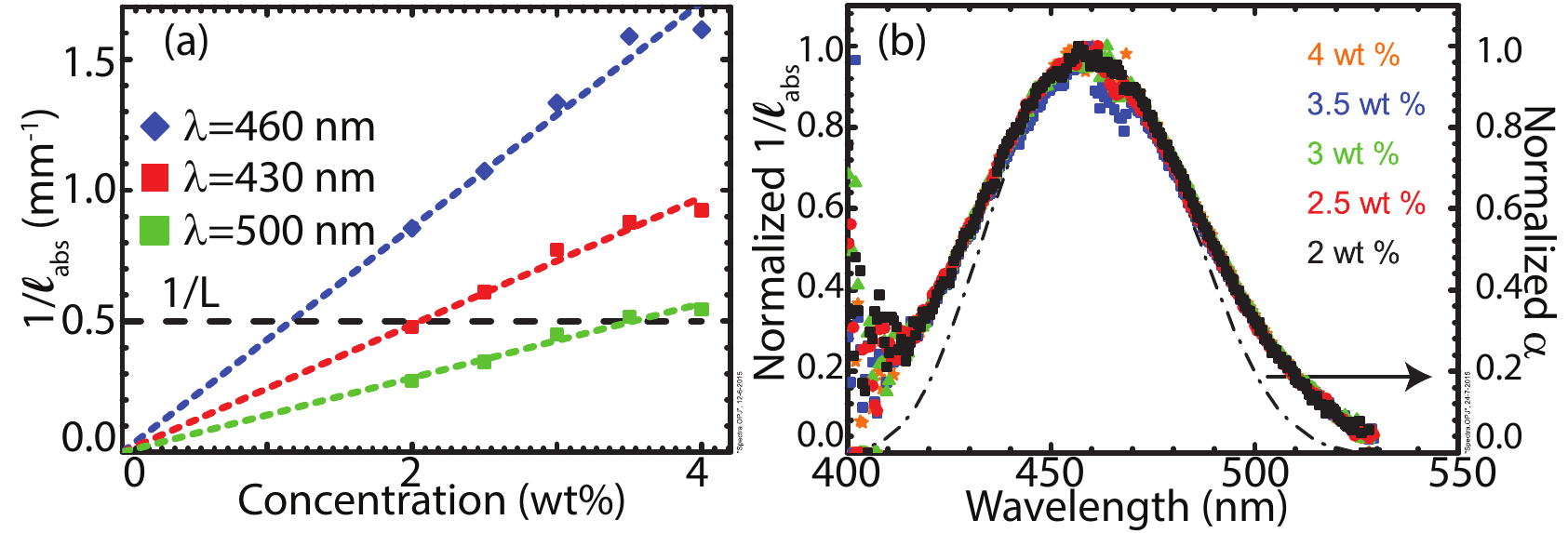}{Absorption mean free path. \textbf{(a)} $\ma$ as a function of concentration is shown for the three different wavelength. \textbf{(b)} Line shapes of the absorption coefficient is shown for the phosphor in powder (dot-dashed black line) and in polymer plates with different phosphor concentrations.}

In Fig.~\ref{fig:trm1}(b) we also compare the shape of the absorption curve of the phosphor in powder form that was used to manufacture the samples to $\ma$ curves. The black dot-dashed line shows absorption spectra of the $\Y$ in powder. These two sets of data were normalized to their maxima, so the positions of maximums of these two graphs coincide. The $\ma$ spectrum appear to have broader tails compared to the $\Y$ in powder form. The absorption spectra of $\Y$ in powder has a FWHM=54 nm, that is 10~nm smaller than the FWHM of the measured absorption spectra. One possible reason is that light is multiply internally reflected in the $\Y$ particles~\cite{Liu10}, so particles can not be treated as an independent point scatterers.

Finally, let us place our approach in the context with previous work: Vos \textit{et} \textit{al.} reported the transport properties measurements in $\mathrm{TiO}_{2}$ scattering plates using a broadband light source~\cite{Vos13}. They showed that the transport mean free path $\lt$ linearly depends on the wavelength in the visible wavelength range. This method can not be applied to the $\Y$ plates, as these plates have ranges with strong absorption, emission and an overlapping range. Leung \textit{et} \textit{al.}~\cite{Leung14} reported the transport properties measurements in $\Y$ plates using a filtered broadband light source, where the linear dependence of $\lt$ was exploited to calculate $\la$. Initially, the approximation used to analyze total transmission $\T$ from Ref.~\cite{Leung14} was used outside its range of validity. Secondly, the described method is limited to the region of strong absorption or emission, but not in the overlap region.  
 
In this paper we have been able to separate light of the same wavelength yet originated from different physical processes occurring in the polymer plates of the solid-state light units. The separate measurement of elastically scattered and Stokes shifted light allowed us to extract the transport and absorption mean free path of the given polymer plates, which are the important parameters required for modeling and predicting the color spectra of solid-state lighting devices. The optimal parameters of the solid-state lighting units can be directly extracted from the measured absorption curves.
We vary the thickness of the plate $L$ or the phosphor concentration depending on the desired level of pump absorption using absorption curve in Fig.~\ref{fig:trm}(b).

\section{\label{ch:Conclusion}Summary and outlook}

We have developed a new technique to measure the light transport of white-light LED plates in the visible range based on narrowband illumination and spectrally sensitive detection. We compare the data obtained with the new technique to the data measure with the broadband light source. The two sets of data coincide in the range without absorption $\lambda\geq\lar$. We extracted the total transmission in the overlap range, that was previously inaccessible.

We used diffusion theory to extract transport $\lt$ and absorption $\la$ mean free paths from these data in the previously inaccessible range. The shape of the absorption coefficient measured for the $\Y$ powder, and $\Y$ powder in polymer matrix have similar trends. Although for the polymer plates the curve is broader then for the $\Y$ powder. Both $\ma$ and $1/\lt$ are proportional to the concentration of phosphor, which reveals that elastic and inelastic processes do not influence each other.
 
By exploiting narrow band light source and interpreting the resulting total transmission by diffusion theory, we are able to extract for the first time light transport parameters for white LEDs in the whole visible wavelength range. However, theory only gives an analytical solution for simple sample geometries, such as a slab, a sphere, or a semi-infinite medium. Therefore, to efficiently model a real white LED with a complex geometry we must in the end supplement an $ab$ $initio$ theory with a numerical method, such as ray-tracing.

\textbf{Acknowledgments}

We would like to thank Cornelis Harteveld for technical support, Vanessa Leung for contribution early on in the project and Teus Tukker, Oluwafemi Ojambati, Ravitej Uppu, Diana Grishina for discussions. This work was supported by the Dutch Technology
Foundation STW (contract no. 11985), and by FOM and NWO, and the ERC (279248).


\footnotetext[1]{In Ref.~\onlinecite{Garcia92} an approximation was used, that is only valid for very small absorption. This approximation lead to the slightly different result in Ref.~\onlinecite{Leung14}. Here we use exact expression and our results are valid for very high $\ma$.}

\bibliography{Narrow_band}

\begin{thebibliography}{41}%
\makeatletter
\providecommand \@ifxundefined [1]{%
 \@ifx{#1\undefined}
}%
\providecommand \@ifnum [1]{%
 \ifnum #1\expandafter \@firstoftwo
 \else \expandafter \@secondoftwo
 \fi
}%
\providecommand \@ifx [1]{%
 \ifx #1\expandafter \@firstoftwo
 \else \expandafter \@secondoftwo
 \fi
}%
\providecommand \natexlab [1]{#1}%
\providecommand \enquote  [1]{``#1''}%
\providecommand \bibnamefont  [1]{#1}%
\providecommand \bibfnamefont [1]{#1}%
\providecommand \citenamefont [1]{#1}%
\providecommand \href@noop [0]{\@secondoftwo}%
\providecommand \href [0]{\begingroup \@sanitize@url \@href}%
\providecommand \@href[1]{\@@startlink{#1}\@@href}%
\providecommand \@@href[1]{\endgroup#1\@@endlink}%
\providecommand \@sanitize@url [0]{\catcode `\\12\catcode `\$12\catcode
  `\&12\catcode `\#12\catcode `\^12\catcode `\_12\catcode `\%12\relax}%
\providecommand \@@startlink[1]{}%
\providecommand \@@endlink[0]{}%
\providecommand \url  [0]{\begingroup\@sanitize@url \@url }%
\providecommand \@url [1]{\endgroup\@href {#1}{\urlprefix }}%
\providecommand \urlprefix  [0]{URL }%
\providecommand \Eprint [0]{\href }%
\providecommand \doibase [0]{http://dx.doi.org/}%
\providecommand \selectlanguage [0]{\@gobble}%
\providecommand \bibinfo  [0]{\@secondoftwo}%
\providecommand \bibfield  [0]{\@secondoftwo}%
\providecommand \translation [1]{[#1]}%
\providecommand \BibitemOpen [0]{}%
\providecommand \bibitemStop [0]{}%
\providecommand \bibitemNoStop [0]{.\EOS\space}%
\providecommand \EOS [0]{\spacefactor3000\relax}%
\providecommand \BibitemShut  [1]{\csname bibitem#1\endcsname}%
\let\auto@bib@innerbib\@empty
\bibitem [{\citenamefont {Schmidt}\ \emph {et~al.}(1999)\citenamefont
  {Schmidt}, \citenamefont {Reichert}, \citenamefont {Ozker}, \citenamefont
  {Meyer}, \citenamefont {Donohoe}, \citenamefont {Bajic}, \citenamefont
  {Whelan},\ and\ \citenamefont {Whelan}}]{Schmidt99}%
  \BibitemOpen
  \bibfield  {author} {\bibinfo {author} {\bibfnamefont {M.~H.}\ \bibnamefont
  {Schmidt}}, \bibinfo {author} {\bibfnamefont {K.~W.}\ \bibnamefont
  {Reichert}}, \bibinfo {author} {\bibfnamefont {K.}~\bibnamefont {Ozker}},
  \bibinfo {author} {\bibfnamefont {G.~A.}\ \bibnamefont {Meyer}}, \bibinfo
  {author} {\bibfnamefont {D.~L.}\ \bibnamefont {Donohoe}}, \bibinfo {author}
  {\bibfnamefont {D.~M.}\ \bibnamefont {Bajic}}, \bibinfo {author}
  {\bibfnamefont {N.~T.}\ \bibnamefont {Whelan}}, \ and\ \bibinfo {author}
  {\bibfnamefont {H.~T.}\ \bibnamefont {Whelan}},\ }\href {\doibase
  10.1159/000028802} {\bibfield  {journal} {\bibinfo  {journal} {Pediatr.
  Neurosurg.}\ }\textbf {\bibinfo {volume} {30}},\ \bibinfo {pages} {225}
  (\bibinfo {year} {1999})}\BibitemShut {NoStop}%
\bibitem [{\citenamefont {Whelan}\ \emph {et~al.}(2000)\citenamefont {Whelan},
  \citenamefont {Houle}, \citenamefont {Whelan}, \citenamefont {Donohoe},
  \citenamefont {Cwiklinski}, \citenamefont {Schmidt}, \citenamefont {Gould},
  \citenamefont {Larson}, \citenamefont {Meyer}, \citenamefont {Cevenini},\
  and\ \citenamefont {Stinson}}]{Whelan00}%
  \BibitemOpen
  \bibfield  {author} {\bibinfo {author} {\bibfnamefont {H.~T.}\ \bibnamefont
  {Whelan}}, \bibinfo {author} {\bibfnamefont {J.~M.}\ \bibnamefont {Houle}},
  \bibinfo {author} {\bibfnamefont {N.~T.}\ \bibnamefont {Whelan}}, \bibinfo
  {author} {\bibfnamefont {D.~L.}\ \bibnamefont {Donohoe}}, \bibinfo {author}
  {\bibfnamefont {J.}~\bibnamefont {Cwiklinski}}, \bibinfo {author}
  {\bibfnamefont {M.~H.}\ \bibnamefont {Schmidt}}, \bibinfo {author}
  {\bibfnamefont {L.}~\bibnamefont {Gould}}, \bibinfo {author} {\bibfnamefont
  {D.~L.}\ \bibnamefont {Larson}}, \bibinfo {author} {\bibfnamefont {G.~A.}\
  \bibnamefont {Meyer}}, \bibinfo {author} {\bibfnamefont {V.}~\bibnamefont
  {Cevenini}}, \ and\ \bibinfo {author} {\bibfnamefont {H.}~\bibnamefont
  {Stinson}},\ }\href {\doibase 10.1063/1.1302454} {\bibfield  {journal}
  {\bibinfo  {journal} {AIP Conference Proceedings}\ }\textbf {\bibinfo
  {volume} {504}},\ \bibinfo {pages} {37} (\bibinfo {year} {2000})}\BibitemShut
  {NoStop}%
\bibitem [{\citenamefont {Malakoff}(2002)}]{Malakoff02}%
  \BibitemOpen
  \bibfield  {author} {\bibinfo {author} {\bibfnamefont {D.}~\bibnamefont
  {Malakoff}},\ }\href {\doibase 10.1126/science.296.5574.1782a} {\bibfield
  {journal} {\bibinfo  {journal} {Science}\ }\textbf {\bibinfo {volume}
  {296}},\ \bibinfo {pages} {1782} (\bibinfo {year} {2002})}\BibitemShut
  {NoStop}%
\bibitem [{\citenamefont {Eells}\ \emph {et~al.}(2004)\citenamefont {Eells},
  \citenamefont {Wong-Riley}, \citenamefont {VerHoeve}, \citenamefont {Henry},
  \citenamefont {Buchman}, \citenamefont {Kane}, \citenamefont {Gould},
  \citenamefont {Das}, \citenamefont {Jett}, \citenamefont {Hodgson},
  \citenamefont {Margolis},\ and\ \citenamefont {Whelan}}]{Eells04}%
  \BibitemOpen
  \bibfield  {author} {\bibinfo {author} {\bibfnamefont {J.~T.}\ \bibnamefont
  {Eells}}, \bibinfo {author} {\bibfnamefont {M.~T.~T.}\ \bibnamefont
  {Wong-Riley}}, \bibinfo {author} {\bibfnamefont {J.}~\bibnamefont
  {VerHoeve}}, \bibinfo {author} {\bibfnamefont {M.}~\bibnamefont {Henry}},
  \bibinfo {author} {\bibfnamefont {E.~V.}\ \bibnamefont {Buchman}}, \bibinfo
  {author} {\bibfnamefont {M.~P.}\ \bibnamefont {Kane}}, \bibinfo {author}
  {\bibfnamefont {L.~J.}\ \bibnamefont {Gould}}, \bibinfo {author}
  {\bibfnamefont {R.}~\bibnamefont {Das}}, \bibinfo {author} {\bibfnamefont
  {M.}~\bibnamefont {Jett}}, \bibinfo {author} {\bibfnamefont {B.~D.}\
  \bibnamefont {Hodgson}}, \bibinfo {author} {\bibfnamefont {D.}~\bibnamefont
  {Margolis}}, \ and\ \bibinfo {author} {\bibfnamefont {H.~T.}\ \bibnamefont
  {Whelan}},\ }\href {\doibase 10.1016/j.mito.2004.07.033} {\bibfield
  {journal} {\bibinfo  {journal} {Mitochondrion}\ }\textbf {\bibinfo {volume}
  {4}},\ \bibinfo {pages} {559 } (\bibinfo {year} {2004})}\BibitemShut
  {NoStop}%
\bibitem [{\citenamefont {Taguchi}, \citenamefont {Uchida},\ and\ \citenamefont
  {Kobashi}(2004)}]{Taguchi04}%
  \BibitemOpen
  \bibfield  {author} {\bibinfo {author} {\bibfnamefont {T.}~\bibnamefont
  {Taguchi}}, \bibinfo {author} {\bibfnamefont {Y.}~\bibnamefont {Uchida}}, \
  and\ \bibinfo {author} {\bibfnamefont {K.}~\bibnamefont {Kobashi}},\ }\href
  {\doibase 10.1002/pssa.200405101} {\bibfield  {journal} {\bibinfo  {journal}
  {Phys. Status Solidi A}\ }\textbf {\bibinfo {volume} {201}},\ \bibinfo
  {pages} {2730} (\bibinfo {year} {2004})}\BibitemShut {NoStop}%
\bibitem [{\citenamefont {Schubert}(2006)}]{Schubert06}%
  \BibitemOpen
  \bibfield  {author} {\bibinfo {author} {\bibfnamefont {E.~F.}\ \bibnamefont
  {Schubert}},\ }\href {http://dx.doi.org/10.1017/CBO9780511790546} {\emph
  {\bibinfo {title} {Light-Emitting Diodes}}},\ \bibinfo {edition} {2nd}\ ed.\
  (\bibinfo  {publisher} {Cambridge University Press},\ \bibinfo {year}
  {2006})\BibitemShut {NoStop}%
\bibitem [{\citenamefont {Krames}\ \emph {et~al.}(2007)\citenamefont {Krames},
  \citenamefont {Shchekin}, \citenamefont {Mueller-Mach}, \citenamefont
  {Mueller}, \citenamefont {Zhou}, \citenamefont {Harbers},\ and\ \citenamefont
  {Craford}}]{Krames07}%
  \BibitemOpen
  \bibfield  {author} {\bibinfo {author} {\bibfnamefont {M.}~\bibnamefont
  {Krames}}, \bibinfo {author} {\bibfnamefont {O.}~\bibnamefont {Shchekin}},
  \bibinfo {author} {\bibfnamefont {R.}~\bibnamefont {Mueller-Mach}}, \bibinfo
  {author} {\bibfnamefont {G.~O.}\ \bibnamefont {Mueller}}, \bibinfo {author}
  {\bibfnamefont {L.}~\bibnamefont {Zhou}}, \bibinfo {author} {\bibfnamefont
  {G.}~\bibnamefont {Harbers}}, \ and\ \bibinfo {author} {\bibfnamefont
  {M.}~\bibnamefont {Craford}},\ }\href {\doibase 10.1109/JDT.2007.895339}
  {\bibfield  {journal} {\bibinfo  {journal} {J. Disp. Tech.,}\ }\textbf
  {\bibinfo {volume} {3}},\ \bibinfo {pages} {160} (\bibinfo {year}
  {2007})}\BibitemShut {NoStop}%
\bibitem [{\citenamefont {Bechtel}\ \emph {et~al.}(2008)\citenamefont
  {Bechtel}, \citenamefont {Schmidt}, \citenamefont {Busselt},\ and\
  \citenamefont {Schreinemacher}}]{Bechtel08}%
  \BibitemOpen
  \bibfield  {author} {\bibinfo {author} {\bibfnamefont {H.}~\bibnamefont
  {Bechtel}}, \bibinfo {author} {\bibfnamefont {P.}~\bibnamefont {Schmidt}},
  \bibinfo {author} {\bibfnamefont {W.}~\bibnamefont {Busselt}}, \ and\
  \bibinfo {author} {\bibfnamefont {B.~S.}\ \bibnamefont {Schreinemacher}},\
  }\href {\doibase 10.1117/12.794941} {\bibfield  {journal} {\bibinfo
  {journal} {Proc. SPIE,}\ }\textbf {\bibinfo {volume} {7058}},\ \bibinfo
  {pages} {70580E} (\bibinfo {year} {2008})}\BibitemShut {NoStop}%
\bibitem [{\citenamefont {Breslauer}\ \emph {et~al.}(2009)\citenamefont
  {Breslauer}, \citenamefont {Maamari}, \citenamefont {Switz}, \citenamefont
  {Lam},\ and\ \citenamefont {Fletcher}}]{Breslauer09}%
  \BibitemOpen
  \bibfield  {author} {\bibinfo {author} {\bibfnamefont {D.~N.}\ \bibnamefont
  {Breslauer}}, \bibinfo {author} {\bibfnamefont {R.~N.}\ \bibnamefont
  {Maamari}}, \bibinfo {author} {\bibfnamefont {N.~A.}\ \bibnamefont {Switz}},
  \bibinfo {author} {\bibfnamefont {W.~A.}\ \bibnamefont {Lam}}, \ and\
  \bibinfo {author} {\bibfnamefont {D.~A.}\ \bibnamefont {Fletcher}},\ }\href
  {\doibase 10.1371/journal.pone.0006320} {\bibfield  {journal} {\bibinfo
  {journal} {PLoS ONE}\ }\textbf {\bibinfo {volume} {4}},\ \bibinfo {pages}
  {e6320} (\bibinfo {year} {2009})}\BibitemShut {NoStop}%
\bibitem [{\citenamefont {Sommer}\ \emph {et~al.}(2009)\citenamefont {Sommer},
  \citenamefont {Krenn}, \citenamefont {Hartmann}, \citenamefont {Pachler},
  \citenamefont {Schweighart}, \citenamefont {Tasch},\ and\ \citenamefont
  {Wenzl}}]{Sommer09}%
  \BibitemOpen
  \bibfield  {author} {\bibinfo {author} {\bibfnamefont {C.}~\bibnamefont
  {Sommer}}, \bibinfo {author} {\bibfnamefont {J.~R.}\ \bibnamefont {Krenn}},
  \bibinfo {author} {\bibfnamefont {P.}~\bibnamefont {Hartmann}}, \bibinfo
  {author} {\bibfnamefont {P.}~\bibnamefont {Pachler}}, \bibinfo {author}
  {\bibfnamefont {M.}~\bibnamefont {Schweighart}}, \bibinfo {author}
  {\bibfnamefont {S.}~\bibnamefont {Tasch}}, \ and\ \bibinfo {author}
  {\bibfnamefont {F.~P.}\ \bibnamefont {Wenzl}},\ }\href {\doibase
  10.1109/JSTQE.2009.2015677} {\bibfield  {journal} {\bibinfo  {journal} {IEEE
  J. Quantum Electron.,}\ }\textbf {\bibinfo {volume} {15}},\ \bibinfo {pages}
  {1181} (\bibinfo {year} {2009})}\BibitemShut {NoStop}%
\bibitem [{\citenamefont {Akasaki}, \citenamefont {Amano},\ and\ \citenamefont
  {Nakamura}(2015)}]{Aka14}%
  \BibitemOpen
  \bibfield  {author} {\bibinfo {author} {\bibfnamefont {I.}~\bibnamefont
  {Akasaki}}, \bibinfo {author} {\bibfnamefont {H.}~\bibnamefont {Amano}}, \
  and\ \bibinfo {author} {\bibfnamefont {S.}~\bibnamefont {Nakamura}},\
  }\href@noop {} {}\bibinfo {howpublished} {See
  \url{http://www.nobelprize.org/}} (\bibinfo {year} {2015})\BibitemShut
  {NoStop}%
\bibitem [{\citenamefont {Gilray}\ and\ \citenamefont
  {Lewin}(1996)}]{Gilray96}%
  \BibitemOpen
  \bibfield  {author} {\bibinfo {author} {\bibfnamefont {C.}~\bibnamefont
  {Gilray}}\ and\ \bibinfo {author} {\bibfnamefont {I.}~\bibnamefont {Lewin}},\
  }\href@noop {} {\bibfield  {journal} {\bibinfo  {journal} {Illuminating
  Engineering Society of North America Annual Conference Technical Papers
  (IESNA). Paper no. 85, pp.}\ ,\ \bibinfo {pages} {65}} (\bibinfo {year}
  {1996})}\BibitemShut {NoStop}%
\bibitem [{\citenamefont {Liu}\ \emph {et~al.}(2010)\citenamefont {Liu},
  \citenamefont {Liu}, \citenamefont {Wang},\ and\ \citenamefont
  {Luo}}]{Liu10}%
  \BibitemOpen
  \bibfield  {author} {\bibinfo {author} {\bibfnamefont {Z.}~\bibnamefont
  {Liu}}, \bibinfo {author} {\bibfnamefont {S.}~\bibnamefont {Liu}}, \bibinfo
  {author} {\bibfnamefont {K.}~\bibnamefont {Wang}}, \ and\ \bibinfo {author}
  {\bibfnamefont {X.}~\bibnamefont {Luo}},\ }\href {\doibase
  10.1364/AO.49.000247} {\bibfield  {journal} {\bibinfo  {journal} {Appl.
  Opt.}\ }\textbf {\bibinfo {volume} {49}},\ \bibinfo {pages} {247} (\bibinfo
  {year} {2010})}\BibitemShut {NoStop}%
\bibitem [{\citenamefont {Tukker}(2010)}]{Tukker10}%
  \BibitemOpen
  \bibfield  {author} {\bibinfo {author} {\bibfnamefont {T.}~\bibnamefont
  {Tukker}},\ }\href@noop {} {\bibfield  {journal} {\bibinfo  {journal} {SPIE
  International Optical Design Conference 2010 (SPIE,2010), Paper no. ITuE2}\ }
  (\bibinfo {year} {2010})}\BibitemShut {NoStop}%
\bibitem [{\citenamefont {Vos}\ \emph {et~al.}(2013)\citenamefont {Vos},
  \citenamefont {Tukker}, \citenamefont {Mosk}, \citenamefont {Lagendijk},\
  and\ \citenamefont {IJzerman}}]{Vos13}%
  \BibitemOpen
  \bibfield  {author} {\bibinfo {author} {\bibfnamefont {W.~L.}\ \bibnamefont
  {Vos}}, \bibinfo {author} {\bibfnamefont {T.~W.}\ \bibnamefont {Tukker}},
  \bibinfo {author} {\bibfnamefont {A.~P.}\ \bibnamefont {Mosk}}, \bibinfo
  {author} {\bibfnamefont {A.}~\bibnamefont {Lagendijk}}, \ and\ \bibinfo
  {author} {\bibfnamefont {W.~L.}\ \bibnamefont {IJzerman}},\ }\href {\doibase
  10.1364/AO.52.002602} {\bibfield  {journal} {\bibinfo  {journal} {Appl.
  Opt.}\ }\textbf {\bibinfo {volume} {52}},\ \bibinfo {pages} {2602} (\bibinfo
  {year} {2013})}\BibitemShut {NoStop}%
\bibitem [{\citenamefont {Leung}\ \emph {et~al.}(2014)\citenamefont {Leung},
  \citenamefont {Lagendijk}, \citenamefont {Tukker}, \citenamefont {Mosk},
  \citenamefont {IJzerman},\ and\ \citenamefont {Vos}}]{Leung14}%
  \BibitemOpen
  \bibfield  {author} {\bibinfo {author} {\bibfnamefont {V.~Y.~F.}\
  \bibnamefont {Leung}}, \bibinfo {author} {\bibfnamefont {A.}~\bibnamefont
  {Lagendijk}}, \bibinfo {author} {\bibfnamefont {T.~W.}\ \bibnamefont
  {Tukker}}, \bibinfo {author} {\bibfnamefont {A.~P.}\ \bibnamefont {Mosk}},
  \bibinfo {author} {\bibfnamefont {W.~L.}\ \bibnamefont {IJzerman}}, \ and\
  \bibinfo {author} {\bibfnamefont {W.~L.}\ \bibnamefont {Vos}},\ }\href
  {\doibase 10.1364/OE.22.008190} {\bibfield  {journal} {\bibinfo  {journal}
  {Opt. Express}\ }\textbf {\bibinfo {volume} {22}},\ \bibinfo {pages} {8190}
  (\bibinfo {year} {2014})}\BibitemShut {NoStop}%
\bibitem [{\citenamefont {Malacara}(2011)}]{Malacara11}%
  \BibitemOpen
  \bibfield  {author} {\bibinfo {author} {\bibfnamefont {D.}~\bibnamefont
  {Malacara}},\ }\href@noop {} {\emph {\bibinfo {title} {Color vision and
  colorimetry: theory and applications}}}\ (\bibinfo  {publisher} {SPIE,
  Washington},\ \bibinfo {year} {2011})\BibitemShut {NoStop}%
\bibitem [{\citenamefont {Lagendijk}\ and\ \citenamefont {van
  Tiggelen}(1996)}]{Lagendijk96}%
  \BibitemOpen
  \bibfield  {author} {\bibinfo {author} {\bibfnamefont {A.}~\bibnamefont
  {Lagendijk}}\ and\ \bibinfo {author} {\bibfnamefont {B.~A.}\ \bibnamefont
  {van Tiggelen}},\ }\href {\doibase
  http://dx.doi.org/10.1016/0370-1573(95)00065-8} {\bibfield  {journal}
  {\bibinfo  {journal} {Phys. Rep.,}\ }\textbf {\bibinfo {volume} {270}},\
  \bibinfo {pages} {143 } (\bibinfo {year} {1996})}\BibitemShut {NoStop}%
\bibitem [{\citenamefont {van Rossum}\ and\ \citenamefont
  {Nieuwenhuizen}(1999)}]{Rossum99}%
  \BibitemOpen
  \bibfield  {author} {\bibinfo {author} {\bibfnamefont {M.~C.~W.}\
  \bibnamefont {van Rossum}}\ and\ \bibinfo {author} {\bibfnamefont {T.~M.}\
  \bibnamefont {Nieuwenhuizen}},\ }\href {\doibase 10.1103/RevModPhys.71.313}
  {\bibfield  {journal} {\bibinfo  {journal} {Rev. Mod. Phys.}\ }\textbf
  {\bibinfo {volume} {71}},\ \bibinfo {pages} {313} (\bibinfo {year}
  {1999})}\BibitemShut {NoStop}%
\bibitem [{\citenamefont {Akkermans}\ and\ \citenamefont
  {Montambaux}(2007)}]{Akkermans07}%
  \BibitemOpen
  \bibfield  {author} {\bibinfo {author} {\bibfnamefont {E.}~\bibnamefont
  {Akkermans}}\ and\ \bibinfo {author} {\bibfnamefont {G.}~\bibnamefont
  {Montambaux}},\ }\href {http://dx.doi.org/10.1017/CBO9780511618833} {\emph
  {\bibinfo {title} {Mesoscopic Physics of Electrons and Photons}}}\ (\bibinfo
  {publisher} {Cambridge University Press},\ \bibinfo {year}
  {2007})\BibitemShut {NoStop}%
\bibitem [{\citenamefont {Garcia}\ \emph {et~al.}(2008)\citenamefont {Garcia},
  \citenamefont {Sapienza}, \citenamefont {Bertolotti}, \citenamefont {Martin},
  \citenamefont {Blanco}, \citenamefont {Altube}, \citenamefont {Vina},
  \citenamefont {Wiersma},\ and\ \citenamefont {Lopez}}]{Garcia08}%
  \BibitemOpen
  \bibfield  {author} {\bibinfo {author} {\bibfnamefont {P.~D.}\ \bibnamefont
  {Garcia}}, \bibinfo {author} {\bibfnamefont {R.}~\bibnamefont {Sapienza}},
  \bibinfo {author} {\bibfnamefont {J.}~\bibnamefont {Bertolotti}}, \bibinfo
  {author} {\bibfnamefont {M.~D.}\ \bibnamefont {Martin}}, \bibinfo {author}
  {\bibfnamefont {A.}~\bibnamefont {Blanco}}, \bibinfo {author} {\bibfnamefont
  {A.}~\bibnamefont {Altube}}, \bibinfo {author} {\bibfnamefont
  {L.}~\bibnamefont {Vina}}, \bibinfo {author} {\bibfnamefont {D.~S.}\
  \bibnamefont {Wiersma}}, \ and\ \bibinfo {author} {\bibfnamefont
  {C.}~\bibnamefont {Lopez}},\ }\href {\doibase 10.1103/PhysRevA.78.023823}
  {\bibfield  {journal} {\bibinfo  {journal} {Phys. Rev. A}\ }\textbf {\bibinfo
  {volume} {78}},\ \bibinfo {pages} {023823} (\bibinfo {year}
  {2008})}\BibitemShut {NoStop}%
\bibitem [{\citenamefont {Muskens}\ and\ \citenamefont
  {Lagendijk}(2008)}]{Muskens08}%
  \BibitemOpen
  \bibfield  {author} {\bibinfo {author} {\bibfnamefont {O.~L.}\ \bibnamefont
  {Muskens}}\ and\ \bibinfo {author} {\bibfnamefont {A.}~\bibnamefont
  {Lagendijk}},\ }\href {\doibase 10.1364/OE.16.001222} {\bibfield  {journal}
  {\bibinfo  {journal} {Opt. Express}\ }\textbf {\bibinfo {volume} {16}},\
  \bibinfo {pages} {1222} (\bibinfo {year} {2008})}\BibitemShut {NoStop}%
\bibitem [{\citenamefont {van~de Hulst}(1980)}]{Hulst80}%
  \BibitemOpen
  \bibfield  {author} {\bibinfo {author} {\bibfnamefont {H.}~\bibnamefont
  {van~de Hulst}},\ }\href@noop {} {\emph {\bibinfo {title} {Multiple Light
  Scattering}}}\ (\bibinfo  {publisher} {Academic Press, Leiden},\ \bibinfo
  {year} {1980})\BibitemShut {NoStop}%
\bibitem [{\citenamefont {Kokhanovsky}(2015)}]{Kokhanovsky2015}%
  \BibitemOpen
  \bibfield  {author} {\bibinfo {author} {\bibfnamefont {A.}~\bibnamefont
  {Kokhanovsky}},\ }\href@noop {} {\emph {\bibinfo {title} {Light Scattering
  Reviews 9}}}\ (\bibinfo  {publisher} {Springer-Verlag, Berlin Heidelberg},\
  \bibinfo {year} {2015})\BibitemShut {NoStop}%
\bibitem [{\citenamefont {Ishimaru}(1978)}]{Ishimaru78}%
  \BibitemOpen
  \bibfield  {author} {\bibinfo {author} {\bibfnamefont {A.}~\bibnamefont
  {Ishimaru}},\ }\href@noop {} {\emph {\bibinfo {title} {Wave propagation and
  scattering in random media}}}\ (\bibinfo  {publisher} {Academic},\ \bibinfo
  {address} {Vols. I and II},\ \bibinfo {year} {1978})\BibitemShut {NoStop}%
\bibitem [{\citenamefont {Garcia}, \citenamefont {Genack},\ and\ \citenamefont
  {Lisyansky}(1992)}]{Garcia92}%
  \BibitemOpen
  \bibfield  {author} {\bibinfo {author} {\bibfnamefont {N.}~\bibnamefont
  {Garcia}}, \bibinfo {author} {\bibfnamefont {A.~Z.}\ \bibnamefont {Genack}},
  \ and\ \bibinfo {author} {\bibfnamefont {A.~A.}\ \bibnamefont {Lisyansky}},\
  }\href {\doibase 10.1103/PhysRevB.46.14475} {\bibfield  {journal} {\bibinfo
  {journal} {Phys. Rev. B}\ }\textbf {\bibinfo {volume} {46}},\ \bibinfo
  {pages} {14475} (\bibinfo {year} {1992})}\BibitemShut {NoStop}%
\bibitem [{\citenamefont {Durian}(1995)}]{Durian95}%
  \BibitemOpen
  \bibfield  {author} {\bibinfo {author} {\bibfnamefont {D.~J.}\ \bibnamefont
  {Durian}},\ }\href {\doibase 10.1364/AO.34.007100} {\bibfield  {journal}
  {\bibinfo  {journal} {Appl. Opt.}\ }\textbf {\bibinfo {volume} {34}},\
  \bibinfo {pages} {7100} (\bibinfo {year} {1995})}\BibitemShut {NoStop}%
\bibitem [{\citenamefont {Elaloufi}, \citenamefont {Carminati},\ and\
  \citenamefont {Greffet}(2002)}]{Elaloufi02}%
  \BibitemOpen
  \bibfield  {author} {\bibinfo {author} {\bibfnamefont {R.}~\bibnamefont
  {Elaloufi}}, \bibinfo {author} {\bibfnamefont {R.}~\bibnamefont {Carminati}},
  \ and\ \bibinfo {author} {\bibfnamefont {J.-J.}\ \bibnamefont {Greffet}},\
  }\href {http://stacks.iop.org/1464-4258/4/i=5/a=355} {\bibfield  {journal}
  {\bibinfo  {journal} {J. Opt. A: Pure Appl. Opt.}\ }\textbf {\bibinfo
  {volume} {4}},\ \bibinfo {pages} {S103} (\bibinfo {year} {2002})}\BibitemShut
  {NoStop}%
\bibitem [{\citenamefont {Sarma}\ \emph {et~al.}(2015)\citenamefont {Sarma},
  \citenamefont {Yamilov}, \citenamefont {Liew}, \citenamefont {Guy},\ and\
  \citenamefont {Cao}}]{Sarma15}%
  \BibitemOpen
  \bibfield  {author} {\bibinfo {author} {\bibfnamefont {R.}~\bibnamefont
  {Sarma}}, \bibinfo {author} {\bibfnamefont {A.}~\bibnamefont {Yamilov}},
  \bibinfo {author} {\bibfnamefont {S.~F.}\ \bibnamefont {Liew}}, \bibinfo
  {author} {\bibfnamefont {M.}~\bibnamefont {Guy}}, \ and\ \bibinfo {author}
  {\bibfnamefont {H.}~\bibnamefont {Cao}},\ }\href@noop {} {\bibfield
  {journal} {\bibinfo  {journal} {arXiv:1507.07861}\ } (\bibinfo {year}
  {2015})}\BibitemShut {NoStop}%
\bibitem [{Note1()}]{Note1}%
  \BibitemOpen
  \bibinfo {note} {In Ref.~\protect \rev@citealpnum {Garcia92} an approximation
  was used, that is only valid for very small absorption. This approximation
  lead to the slightly different result in Ref.~\protect \rev@citealpnum
  {Leung14}. Here we use exact expression and our results are valid for very
  high $\mu _{\protect \mathrm {a}}$.}\BibitemShut {Stop}%
\bibitem [{\citenamefont {Lagendijk}, \citenamefont {Vreeker},\ and\
  \citenamefont {Vries}(1989)}]{Lagendijk89}%
  \BibitemOpen
  \bibfield  {author} {\bibinfo {author} {\bibfnamefont {A.}~\bibnamefont
  {Lagendijk}}, \bibinfo {author} {\bibfnamefont {R.}~\bibnamefont {Vreeker}},
  \ and\ \bibinfo {author} {\bibfnamefont {P.~D.}\ \bibnamefont {Vries}},\
  }\href {\doibase http://dx.doi.org/10.1016/0375-9601(89)90683-X} {\bibfield
  {journal} {\bibinfo  {journal} {Phys. Lett. A}\ }\textbf {\bibinfo {volume}
  {136}},\ \bibinfo {pages} {81 } (\bibinfo {year} {1989})}\BibitemShut
  {NoStop}%
\bibitem [{\citenamefont {Rivas}\ \emph {et~al.}(1999)\citenamefont {Rivas},
  \citenamefont {Sprik}, \citenamefont {Soukoulis}, \citenamefont {Busch},\
  and\ \citenamefont {Lagendijk}}]{Gomez99}%
  \BibitemOpen
  \bibfield  {author} {\bibinfo {author} {\bibfnamefont {J.~G.}\ \bibnamefont
  {Rivas}}, \bibinfo {author} {\bibfnamefont {R.}~\bibnamefont {Sprik}},
  \bibinfo {author} {\bibfnamefont {C.~M.}\ \bibnamefont {Soukoulis}}, \bibinfo
  {author} {\bibfnamefont {K.}~\bibnamefont {Busch}}, \ and\ \bibinfo {author}
  {\bibfnamefont {A.}~\bibnamefont {Lagendijk}},\ }\href
  {http://stacks.iop.org/0295-5075/48/i=1/a=022} {\bibfield  {journal}
  {\bibinfo  {journal} {Europhys. Lett.}\ }\textbf {\bibinfo {volume} {48}},\
  \bibinfo {pages} {22} (\bibinfo {year} {1999})}\BibitemShut {NoStop}%
\bibitem [{\citenamefont {Zhu}, \citenamefont {Pine},\ and\ \citenamefont
  {Weitz}(1991)}]{Zhu91}%
  \BibitemOpen
  \bibfield  {author} {\bibinfo {author} {\bibfnamefont {J.~X.}\ \bibnamefont
  {Zhu}}, \bibinfo {author} {\bibfnamefont {D.~J.}\ \bibnamefont {Pine}}, \
  and\ \bibinfo {author} {\bibfnamefont {D.~A.}\ \bibnamefont {Weitz}},\ }\href
  {\doibase 10.1103/PhysRevA.44.3948} {\bibfield  {journal} {\bibinfo
  {journal} {Phys. Rev. A.}\ }\textbf {\bibinfo {volume} {44}},\ \bibinfo
  {pages} {3948} (\bibinfo {year} {1991})}\BibitemShut {NoStop}%
\bibitem [{\citenamefont {Muskens}\ and\ \citenamefont
  {Lagendijk}(2009)}]{Muskens09}%
  \BibitemOpen
  \bibfield  {author} {\bibinfo {author} {\bibfnamefont {O.~L.}\ \bibnamefont
  {Muskens}}\ and\ \bibinfo {author} {\bibfnamefont {A.}~\bibnamefont
  {Lagendijk}},\ }\href {\doibase 10.1364/OL.34.000395} {\bibfield  {journal}
  {\bibinfo  {journal} {Opt. Lett.}\ }\textbf {\bibinfo {volume} {34}},\
  \bibinfo {pages} {395} (\bibinfo {year} {2009})}\BibitemShut {NoStop}%
\bibitem [{LED(2014)}]{LED14}%
  \BibitemOpen
  \href@noop {} {\enquote {\bibinfo {title}
  {See~catalog~at:~\url{http://www.lighting.philips.co.uk/pwc_li/gb_en/subsites/oem/fortimo-led-catalogue}},}\
  } (\bibinfo {year} {2014})\BibitemShut {NoStop}%
\bibitem [{\citenamefont {Busch}, \citenamefont {Soukoulis},\ and\
  \citenamefont {Economou}(1994)}]{Busch94}%
  \BibitemOpen
  \bibfield  {author} {\bibinfo {author} {\bibfnamefont {K.}~\bibnamefont
  {Busch}}, \bibinfo {author} {\bibfnamefont {C.~M.}\ \bibnamefont
  {Soukoulis}}, \ and\ \bibinfo {author} {\bibfnamefont {E.~N.}\ \bibnamefont
  {Economou}},\ }\href {\doibase 10.1103/PhysRevB.50.93} {\bibfield  {journal}
  {\bibinfo  {journal} {Phys. Rev. B}\ }\textbf {\bibinfo {volume} {50}},\
  \bibinfo {pages} {93} (\bibinfo {year} {1994})}\BibitemShut {NoStop}%
\bibitem [{\citenamefont {Kaczmarek}\ \emph {et~al.}(1999)\citenamefont
  {Kaczmarek}, \citenamefont {Domianiak-Dzik}, \citenamefont {Ryba-Romanowski},
  \citenamefont {Kisielewski},\ and\ \citenamefont {Wojtkowska}}]{Kaczmarek99}%
  \BibitemOpen
  \bibfield  {author} {\bibinfo {author} {\bibfnamefont {S.~M.}\ \bibnamefont
  {Kaczmarek}}, \bibinfo {author} {\bibfnamefont {G.}~\bibnamefont
  {Domianiak-Dzik}}, \bibinfo {author} {\bibfnamefont {W.}~\bibnamefont
  {Ryba-Romanowski}}, \bibinfo {author} {\bibfnamefont {J.}~\bibnamefont
  {Kisielewski}}, \ and\ \bibinfo {author} {\bibfnamefont {J.}~\bibnamefont
  {Wojtkowska}},\ }\href {\doibase
  10.1002/(SICI)1521-4079(199909)34:8<1031::AID-CRAT1031>3.0.CO;2-I} {\bibfield
   {journal} {\bibinfo  {journal} {Cryst. Res. Technol.}\ }\textbf {\bibinfo
  {volume} {34}},\ \bibinfo {pages} {1031} (\bibinfo {year}
  {1999})}\BibitemShut {NoStop}%
\bibitem [{\citenamefont {Zhao}\ \emph {et~al.}(2003)\citenamefont {Zhao},
  \citenamefont {Zeng}, \citenamefont {Xu}, \citenamefont {Zhou},\ and\
  \citenamefont {Zhou}}]{Zhao03}%
  \BibitemOpen
  \bibfield  {author} {\bibinfo {author} {\bibfnamefont {G.~J.}\ \bibnamefont
  {Zhao}}, \bibinfo {author} {\bibfnamefont {X.~H.}\ \bibnamefont {Zeng}},
  \bibinfo {author} {\bibfnamefont {J.}~\bibnamefont {Xu}}, \bibinfo {author}
  {\bibfnamefont {S.~M.}\ \bibnamefont {Zhou}}, \ and\ \bibinfo {author}
  {\bibfnamefont {Y.~Z.}\ \bibnamefont {Zhou}},\ }\href {\doibase
  10.1002/pssa.200306652} {\bibfield  {journal} {\bibinfo  {journal} {Phys.
  Status Solidi A}\ }\textbf {\bibinfo {volume} {199}},\ \bibinfo {pages} {355}
  (\bibinfo {year} {2003})}\BibitemShut {NoStop}%
\bibitem [{\citenamefont {Mares}\ \emph {et~al.}(2007)\citenamefont {Mares},
  \citenamefont {Beitlerova}, \citenamefont {Nikl}, \citenamefont {Solovieva},
  \citenamefont {Nitsch}, \citenamefont {Kucera}, \citenamefont {Kubova},
  \citenamefont {Gorbenko},\ and\ \citenamefont {Zorenko}}]{Mares07}%
  \BibitemOpen
  \bibfield  {author} {\bibinfo {author} {\bibfnamefont {J.~A.}\ \bibnamefont
  {Mares}}, \bibinfo {author} {\bibfnamefont {A.}~\bibnamefont {Beitlerova}},
  \bibinfo {author} {\bibfnamefont {M.}~\bibnamefont {Nikl}}, \bibinfo {author}
  {\bibfnamefont {N.}~\bibnamefont {Solovieva}}, \bibinfo {author}
  {\bibfnamefont {K.}~\bibnamefont {Nitsch}}, \bibinfo {author} {\bibfnamefont
  {M.}~\bibnamefont {Kucera}}, \bibinfo {author} {\bibfnamefont
  {M.}~\bibnamefont {Kubova}}, \bibinfo {author} {\bibfnamefont
  {V.}~\bibnamefont {Gorbenko}}, \ and\ \bibinfo {author} {\bibfnamefont
  {Y.}~\bibnamefont {Zorenko}},\ }\href@noop {} {\bibfield  {journal} {\bibinfo
   {journal} {Radiat. Meas.}\ }\textbf {\bibinfo {volume} {42}},\ \bibinfo
  {pages} {533 } (\bibinfo {year} {2007})}\BibitemShut {NoStop}%
\bibitem [{\citenamefont {Mihóková}\ \emph {et~al.}(2007)\citenamefont
  {Mihóková}, \citenamefont {Nikl}, \citenamefont {Mareš}, \citenamefont
  {Beitlerová}, \citenamefont {Vedda}, \citenamefont {Nejezchleb},
  \citenamefont {Blažek},\ and\ \citenamefont {D’Ambrosio}}]{Mihokova07}%
  \BibitemOpen
  \bibfield  {author} {\bibinfo {author} {\bibfnamefont {E.}~\bibnamefont
  {Mihóková}}, \bibinfo {author} {\bibfnamefont {M.}~\bibnamefont {Nikl}},
  \bibinfo {author} {\bibfnamefont {J.}~\bibnamefont {Mareš}}, \bibinfo
  {author} {\bibfnamefont {A.}~\bibnamefont {Beitlerová}}, \bibinfo {author}
  {\bibfnamefont {A.}~\bibnamefont {Vedda}}, \bibinfo {author} {\bibfnamefont
  {K.}~\bibnamefont {Nejezchleb}}, \bibinfo {author} {\bibfnamefont
  {K.}~\bibnamefont {Blažek}}, \ and\ \bibinfo {author} {\bibfnamefont
  {C.}~\bibnamefont {D’Ambrosio}},\ }\href {\doibase
  http://dx.doi.org/10.1016/j.jlumin.2006.05.004} {\bibfield  {journal}
  {\bibinfo  {journal} {J. Lumin.}\ }\textbf {\bibinfo {volume} {126}},\
  \bibinfo {pages} {77 } (\bibinfo {year} {2007})}\BibitemShut {NoStop}%
\bibitem [{\citenamefont {Kučera}, \citenamefont {Hasa},\ and\ \citenamefont
  {Hakenová}(2008)}]{Kucera08}%
  \BibitemOpen
  \bibfield  {author} {\bibinfo {author} {\bibfnamefont {M.}~\bibnamefont
  {Kučera}}, \bibinfo {author} {\bibfnamefont {P.}~\bibnamefont {Hasa}}, \
  and\ \bibinfo {author} {\bibfnamefont {J.}~\bibnamefont {Hakenová}},\
  }\href@noop {} {\bibfield  {journal} {\bibinfo  {journal} {J. Alloy Compd.}\
  }\textbf {\bibinfo {volume} {451}},\ \bibinfo {pages} {146 } (\bibinfo {year}
  {2008})}\BibitemShut {NoStop}%
\end{thebibliography}%

\end{document}